\documentclass[conference,10pt]{IEEEtran}
\IEEEoverridecommandlockouts
\usepackage{cite}
\usepackage{amsmath,amssymb,amsfonts}
\usepackage[nolist]{acronym}
\usepackage[bookmarks=false]{hyperref}
\usepackage{scalerel}
\usepackage{graphicx}
\usepackage{subcaption}
\usepackage{nicefrac}
\usepackage{xcolor}
\usepackage{circuitikz}
\usepackage{adjustbox}
\usepackage{latexml}

\usetikzlibrary{calc}
\usepackage{pgfplots}
\usepgfplotslibrary{groupplots}

\pgfplotsset{compat=1.15}

\usetikzlibrary{positioning}
\usetikzlibrary{svg.path}
\usetikzlibrary{shapes}

\tikzset{
    iso/.style={kite, kite vertex angles=120,  minimum size=0.06cm, inner sep=0.1pt, outer sep=0pt, font = \footnotesize, gray} 
}

\DeclareRobustCommand{\captionslantantenna}[1]{%
    \tikz[baseline=-0.6ex, x=1ex, y=1ex]{
        \draw[line width=0.7pt, magenta] (-0.5,-0.5) -- (0.5,0.5);
        \draw[line width=0.7pt, teal] (0.5,-0.5) -- (-0.5,0.5);
        \draw[dash pattern=on 0.5pt off 0.35pt, line width=0.4pt] (0,0) circle [radius=0.85];
    }%
}

\def\BibTeX{{\rm B\kern-.05em{\sc i\kern-.025em b}\kern-.08em
    T\kern-.1667em\lower.7ex\hbox{E}\kern-.125emX}}

\definecolor{orcidlogocol}{HTML}{A6CE39}
\tikzset{
  orcidlogo/.pic={
    \fill[orcidlogocol] svg{M256,128c0,70.7-57.3,128-128,128C57.3,256,0,198.7,0,128C0,57.3,57.3,0,128,0C198.7,0,256,57.3,256,128z};
    \fill[white] svg{M86.3,186.2H70.9V79.1h15.4v48.4V186.2z}
                 svg{M108.9,79.1h41.6c39.6,0,57,28.3,57,53.6c0,27.5-21.5,53.6-56.8,53.6h-41.8V79.1z M124.3,172.4h24.5c34.9,0,42.9-26.5,42.9-39.7c0-21.5-13.7-39.7-43.7-39.7h-23.7V172.4z}
                 svg{M88.7,56.8c0,5.5-4.5,10.1-10.1,10.1c-5.6,0-10.1-4.6-10.1-10.1c0-5.6,4.5-10.1,10.1-10.1C84.2,46.7,88.7,51.3,88.7,56.8z};
  }
}

\newcommand\orcidicon[1]{\href{https://orcid.org/#1}{\mbox{\scalerel*{
\begin{tikzpicture}[yscale=-1,transform shape]
\pic{orcidlogo};
\end{tikzpicture}
}{|}}}}

\acrodef{CSI}{channel state information}
\acrodef{RF}{radio frequency}
\acrodef{MUSIC}{MUltiple SIgnal Classification}
\acrodef{FFT}{fast Fourier transform}
\acrodef{OFDM}{orthogonal frequency division multiplex}
\acrodef{SNR}{signal-to-noise ratio}
\acrodef{CFO}{carrier frequency offset}
\acrodef{CPU}{central processing unit}
\acrodef{GPU}{graphics processing unit}

\definecolor{mittelblau}{RGB}{0, 126, 198}
\definecolor{violettblau}{cmyk}{0.9, 0.6, 0, 0}
\definecolor{rot}{RGB}{238, 28 35}
\definecolor{apfelgruen}{RGB}{140, 198, 62}
\definecolor{gelb}{RGB}{1, 221, 0}
\definecolor{orange}{RGB}{244, 111, 33}
\definecolor{pink}{RGB}{237, 0, 140}
\definecolor{lila}{RGB}{128, 10, 145}
\definecolor{hellgrau}{RGB}{224, 224, 224}
\definecolor{mittelgrau}{RGB}{128, 128, 128}
\definecolor{dunkelgrau}{RGB}{80,80,80}
\definecolor{anthrazit}{RGB}{19, 31, 31}

\begin{document}

\title{Visualizing Wireless Propagation and Polarization in Augmented Reality with ESPARGOS
\thanks{This work is supported by the German Federal Ministry of Research, Technology and Space (BMFTR) within the SENSATION project (grant no. 16KIS2532).}}

\author{\IEEEauthorblockN{Florian Euchner\textsuperscript{\orcidicon{0000-0002-8090-1188}}, Stephan ten Brink\textsuperscript{\orcidicon{0000-0003-1502-2571}} \\}
\IEEEauthorblockA{
Institute of Telecommunications, Pfaffenwaldring 47, University of Stuttgart, 70569 Stuttgart, Germany \\ \{euchner,tenbrink\}@inue.uni-stuttgart.de
}
}

\newpage

\maketitle

\begin{abstract}
Wireless multipath propagation, beamforming, and polarization are central concepts in radio systems, but they are difficult to observe directly because radio-frequency fields are invisible to humans.
This paper presents an augmented-reality visualization system that turns phase-coherent WiFi channel measurements from the ESPARGOS antenna array into a live camera overlay.
The system estimates a two-dimensional beamspace representation, registers it with the optical camera view, and overlays the resulting angular spectrum onto the physical scene.
Additional visual layers show relative path delay and polarization, the latter computed using Jones calculus from channel measurements captured from two separate antenna feeds.
The result is an intuitive ``WiFi camera'' for science communication, teaching, and experimental debugging, while all displayed quantities remain directly derived from measured channel data.
\end{abstract}

\section{Introduction}
Coherent multi-antenna channel measurements provide a detailed, multi-dimensional view of the physical radio environment.
They reveal where signal energy arrives from, how signals are delayed by reflections, and how polarization changes along the propagation path.
Despite this rich information content, such measurements are usually inspected through abstract plots or processed by communication or localization algorithms.
This makes them difficult to interpret for non-specialists and cumbersome to make sense of even for researchers who want to quickly debug an experiment, demonstrate a propagation effect, or reason about the physical meaning of measured \ac{CSI}.

Previously, we introduced ESPARGOS as an ultra-low-cost, real-time-capable, phase-coherent multi-antenna WiFi channel sounder for wireless sensing experiments~\cite{espargos}.
Subsequent work used ESPARGOS to record public \ac{CSI} datasets~\cite{euchner2024kh} and to demonstrate passive target localization with distributed WiFi sensing setups~\cite{euchner2025spawc}.
This paper follows a complementary direction: instead of using \ac{CSI} only as input to an offline algorithm, we use it as the basis for a live, camera-registered visualization of wireless propagation.
The resulting system can be understood as an augmented-reality layer for channel sounding.
A conventional camera provides the optical scene, while an ESPARGOS array provides a view of the same scene in the 2.4 GHz WiFi band of the \ac{RF} spectrum.
By aligning both views, received power, multipath delay, and polarization can be displayed directly at the apparent directions of the corresponding sources or scatterers.
Prior immersive and augmented-reality systems visualize predicted radio maps, signal strength, or network-planning metrics~\cite{koutitas2020situ,rowden2022waverider,okubo2023field}.
In contrast, our system overlays spatial information about the \ac{RF} signal itself, exposing angular structure, delay, and polarization in a camera-registered view.
The main contribution of this paper is the compact signal-processing and rendering pipeline that connects optical and \ac{RF} views in real time.

\section{System and Visualization Pipeline}
\label{sec:pipeline}

This section describes the processing chain from phase-coherent WiFi \ac{CSI} to an augmented-reality overlay.
Our open-source implementation\footnote{Available from \url{https://github.com/ESPARGOS/pyespargos}} also supports other spatial-spectrum estimators, such as \ac{MUSIC}~\cite{schmidt1986music}, but we focus on the \ac{FFT}-based beamspace method because it is simple, fast, and hence well suited for real-time visualization.

\subsection{ESPARGOS Receiver Setup}

\begin{figure}
    \centering
    \includegraphics[width=0.55\columnwidth]{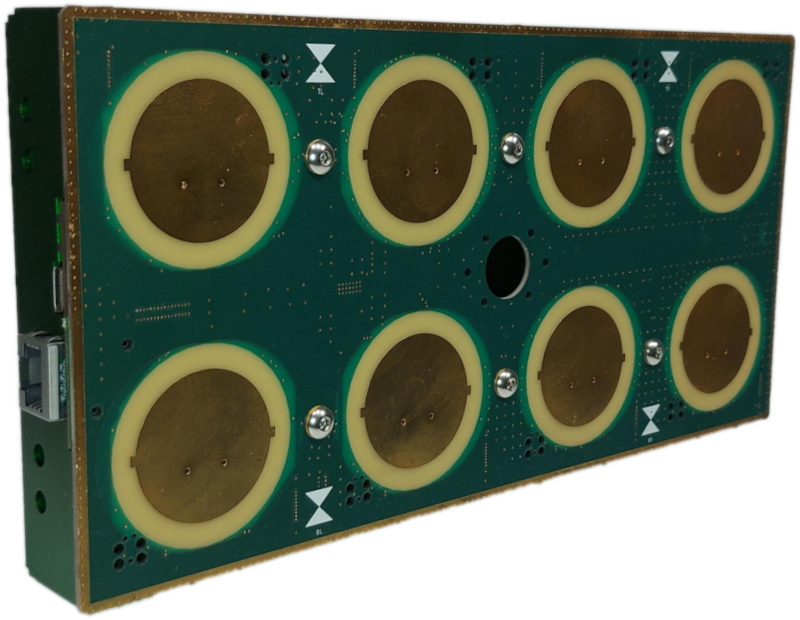}
    \caption{``ESPARGOS One'' antenna array: WiFi-based fully digital phased array with dual slant-polarized patch antennas.}
    \label{fig:espargos-array}
\end{figure}

\begin{figure}
    \centering
    \iflatexml
        \includegraphics[width=\columnwidth]{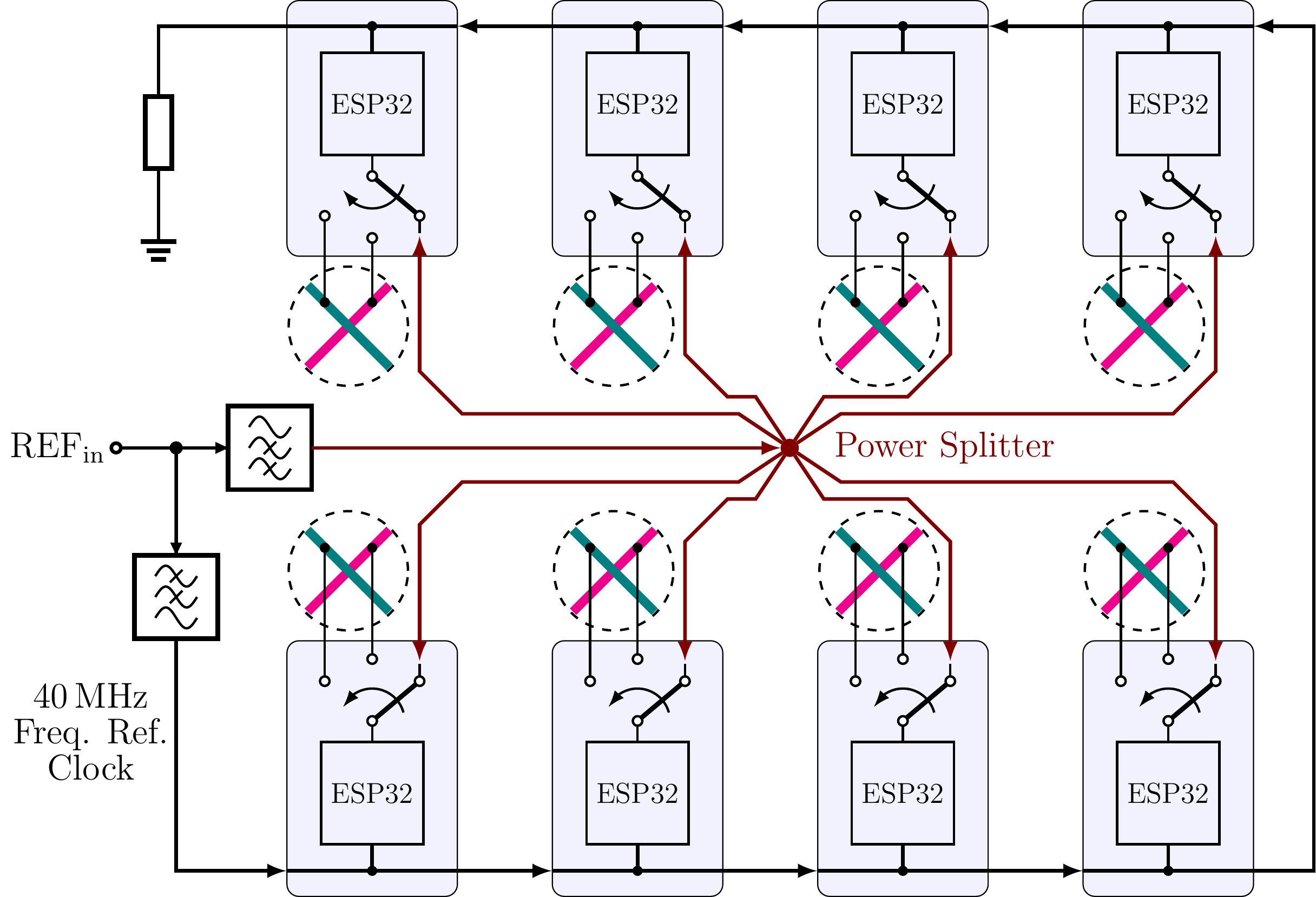}
    \else
        \begin{adjustbox}{max width=\columnwidth}
            \begin{circuitikz}[european]
    \node (ant0) [draw, minimum width = 2cm, minimum height = 3cm, rounded corners, fill = blue!5!white] at (0, 0) {};
    \node (ant1) [draw, minimum width = 2cm, minimum height = 3cm, rounded corners, fill = blue!5!white, right = 1.1cm of ant0] {};
    \node (ant2) [draw, minimum width = 2cm, minimum height = 3cm, rounded corners, fill = blue!5!white, right = 1.1cm of ant1] {};
    \node (ant3) [draw, minimum width = 2cm, minimum height = 3cm, rounded corners, fill = blue!5!white, right = 1.1cm of ant2] {};
    \node (ant4) [draw, minimum width = 2cm, minimum height = 3cm, rounded corners, fill = blue!5!white, below = 4.5cm of ant3] {};
    \node (ant5) [draw, minimum width = 2cm, minimum height = 3cm, rounded corners, fill = blue!5!white, left = 1.1cm of ant4] {};
    \node (ant6) [draw, minimum width = 2cm, minimum height = 3cm, rounded corners, fill = blue!5!white, left = 1.1cm of ant5] {};
    \node (ant7) [draw, minimum width = 2cm, minimum height = 3cm, rounded corners, fill = blue!5!white, left = 1.1cm of ant6] {};

    \draw [very thick, latex-] ($(ant0.east) + (0, 1.2)$) -- ($(ant1.west) + (0, 1.2)$);
    \draw [very thick, latex-] ($(ant1.east) + (0, 1.2)$) -- ($(ant2.west) + (0, 1.2)$);
    \draw [very thick, latex-] ($(ant2.east) + (0, 1.2)$) -- ($(ant3.west) + (0, 1.2)$);
    \draw [very thick, latex-] ($(ant3.east) + (0, 1.2)$) -- +(0.7, 0) |- ($(ant4.east) + (0, -1.2)$);
    \draw [very thick, latex-] ($(ant4.west) + (0, -1.2)$) -- ($(ant5.east) + (0, -1.2)$);
    \draw [very thick, latex-] ($(ant5.west) + (0, -1.2)$) -- ($(ant6.east) + (0, -1.2)$);
    \draw [very thick, latex-] ($(ant6.west) + (0, -1.2)$) -- ($(ant7.east) + (0, -1.2)$);
    
    \draw [very thick] ($(ant0.west) + (0, 1.2)$) -- ++(-1.5, 0) to[/tikz/circuitikz/bipoles/length=30pt, line width = 0.8, R] ++(0, -2.5) node[thick, ground, yscale = 1.5, yshift = 0.4cm] {};

    \newcommand{\bb}[1]{
        \draw [very thick] ($(ant#1.west) + (0, 1.2)$) -- ($(ant#1.east) + (0, 1.2)$);
        \node (esp#1) [scale = 0.8, thick, draw, minimum width = 1.5cm, minimum height = 1.5cm, anchor = north] at ($(ant#1.center) + (0, 0.9)$) {\large ESP32};
        \draw [very thick] ($(ant#1.center) + (0, 1.2)$) node[circle, fill = black, inner sep = 1.2pt] {} -- (esp#1.north);
        \ctikzset{multipoles/rotary/thickness=0.5}
		\node [rotary switch - = 3 in 50 wiper 50, rotate = -90, thick] (rfsw#1) at ($(esp#1.south) + (0, -0.6)$) {};
        \draw [thick] (esp#1.south) -- (rfsw#1.in);
        \draw [thick, -{Latex[length = 2mm, width = 1.4mm]}]
            ([shift={(-15:0.38cm)}]rfsw#1.ext center)
            arc[start angle = -15, end angle = -155, radius = 0.38cm];
        \draw [thick] (rfsw#1.out 1) -- ($(rfsw#1.out 1) + (0, -0.15)$) node [inner sep = 0pt] (refinput#1) {};

        \draw [line width = 3pt, magenta] ($(rfsw#1.out 2) + (0.2, -0.5)$) -- ++(-0.96, -0.96);
        \draw [line width = 3pt, teal] ($(rfsw#1.out 2 -| rfsw#1.out 3) + (-0.2, -0.5)$) -- ++(0.96, -0.96);
        \node [circle, dashed, draw = black, minimum size = 1.4cm, thick, inner sep = 0pt] at ($(rfsw#1.out 2) + (0.2, -0.5) + (-0.48, -0.48)$) {};

        \draw [thick] (rfsw#1.out 2) -- ($(rfsw#1.out 2) + (0, -0.7)$) node[circle, fill = black, inner sep = 1.2pt] {};
        \draw [thick] (rfsw#1.out 3) -- ($(rfsw#1.out 3 |- rfsw#1.out 2) + (0, -0.7)$) node[circle, fill = black, inner sep = 1.2pt] {};
    }

    \newcommand{\bbflip}[1]{
        \draw [very thick] ($(ant#1.west) + (0, -1.2)$) -- ($(ant#1.east) + (0, -1.2)$);
        \node (esp#1) [scale = 0.8, thick, draw, minimum width = 1.5cm, minimum height = 1.5cm, anchor = south] at ($(ant#1.center) + (0, -0.9)$) {\large ESP32};
        \draw [very thick] ($(ant#1.center) + (0, -1.2)$) node[circle, fill = black, inner sep = 1.2pt] {} -- (esp#1.south);
        \begin{scope}[shift={($(esp#1.north) + (0, 0.6)$)}, yscale = -1]
				\node [rotary switch - = 3 in 50 wiper 50, rotate = -90, transform shape, thick] (rfsw#1) at (0, 0) {};
            \draw [thick, -{Latex[length = 2mm, width = 1.4mm]}]
                ([shift={(-15:0.38cm)}]rfsw#1.ext center)
                arc[start angle = -15, end angle = -155, radius = 0.38cm];
            \draw [thick] (rfsw#1.out 1) -- ($(rfsw#1.out 1) + (0, -0.15)$) node [inner sep = 0pt] (refinput#1) {};

            \draw [line width = 3pt, magenta] ($(rfsw#1.out 2 -| rfsw#1.out 3) + (-0.2, -0.5)$) -- ++(0.96, -0.96);
            \draw [line width = 3pt, teal] ($(rfsw#1.out 2) + (0.2, -0.5)$) -- ++(-0.96, -0.96);
            \node [circle, dashed, draw = black, minimum size = 1.4cm, thick, inner sep = 0pt] at ($(rfsw#1.out 2) + (0.2, -0.5) + (-0.48, -0.48)$) {};

            \draw [thick] (rfsw#1.out 2) -- ($(rfsw#1.out 2) + (0, -1.25)$) node[circle, fill = black, inner sep = 1.2pt] {};
            \draw [thick] (rfsw#1.out 3) -- ($(rfsw#1.out 3 |- rfsw#1.out 2) + (0, -1.25)$) node[circle, fill = black, inner sep = 1.2pt] {};
        \end{scope}
        \draw [thick] (esp#1.north) -- (rfsw#1.in);
    }
    
    \bb{0}
    \bb{1}
    \bb{2}
    \bb{3}
    \bbflip{4}
    \bbflip{5}
    \bbflip{6}
    \bbflip{7}

    \newcommand{\phasepath}[4]{
        \node [inner sep = 0pt] (waypointA#1) at ($(refsrc) + (-0.5 * #3 + #2 * #3 * -0.1, 0.5 * #4 + #2 * #4 * -0.1)$) {};
        \node [inner sep = 0pt] (waypointB#1) at (refinput#1 |- waypointA#1) {};
        \draw [red!50!black, very thick, -latex] (refsrc) -- (waypointA#1.center) -- ($(waypointB#1) + (0.5 * #3, 0)$) -- ($(waypointB#1) + (0, 0.5 * #4)$) -- (refinput#1);
    }

    \node [circle, inner sep = 2pt, draw = red!50!black, fill = red!50!black] (refsrc) at (4.9, -3.75) {};
    \node [anchor = west, red!50!black] at ($(refsrc) + (0.4, 0)$) {\large Power Splitter};
	\phasepath{0}{1}{1}{1}
	\phasepath{1}{-1}{1}{1}
	\phasepath{2}{-1}{-1}{1}
	\phasepath{3}{1}{-1}{1}
	\phasepath{7}{1}{1}{-1}
	\phasepath{6}{-1}{1}{-1}
	\phasepath{5}{-1}{-1}{-1}
	\phasepath{4}{1}{-1}{-1}

	\node (hpf) [highpassshape, thick] at (-1.2, -3.75) {};
	\node (lpf) [lowpassshape, thick] at (-2.3, -5.5) {};
	\draw [red!50!black, very thick, -latex] (hpf.east) -- (refsrc);
	\draw [very thick, -latex] (lpf.south) |- ($(ant7.west) + (0, -1.2)$) node[pos = 0.2, align = center, xshift = -1cm] {\large $40\,\mathrm{MHz}$\\\large Freq. Ref.\\\large Clock};
	\node (refin) at ($(hpf.west) + (-2, 0)$) {\large REF\textsubscript{in}};
	\draw [very thick] (refin.east) to[short, o-] (hpf.west) node[inputarrow] {};
	\draw [very thick] (refin -| lpf.north) to[short, *-] (lpf.north) node[inputarrow, rotate = -90] {};
\end{circuitikz}
        \end{adjustbox}
    \fi
    \caption{Block diagram of a single $2\times4$ ESPARGOS antenna array, with a focus on the synchronization and antenna signals. The \protect\captionslantantenna{} symbols denote the antenna patches with two feeds for $\pm 45^\circ$ slant polarization.}
    \label{fig:block-diagram}
\end{figure}
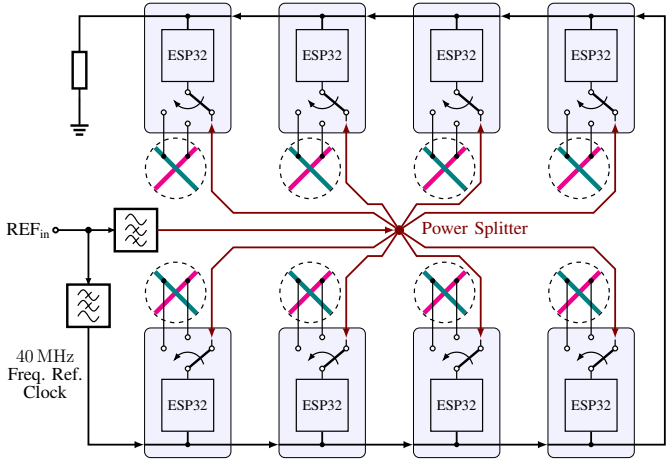

ESPARGOS is used here as a receive-only, phase-coherent WiFi sensing array.
The current hardware generation, shown in Fig.~\ref{fig:espargos-array}, consists of a $2\times4$ patch antenna array with two switchable feeds per element for $\pm 45^\circ$ slant polarization, and operates in the $2.4\,\mathrm{GHz}$ WiFi band.
The WiFi receivers are synchronized using a shared clock and phase reference distribution network, as shown in Fig.~\ref{fig:block-diagram}.
Larger apertures are formed by combining several such subarrays into a single rectangular array.
The software configuration specifies how the antenna positions of the individual boards map to the rows and columns of the combined array.
The raw \ac{CSI} is corrected using the receiver gain values reported by the WiFi chips, phase-calibrated across receivers, and arranged according to the configured array mapping.
The resulting measured frequency-domain channel is denoted by
\begin{equation*}
    H_{m,n,k} \in \mathbb{C},
    \quad
    m=0,\ldots,M-1,\quad n=0,\ldots,N-1,
\end{equation*}
where $M$ and $N$ are the numbers of array rows and columns, $(m,n)$ indexes the antenna element, and $k$ indexes the \ac{OFDM} subcarrier at frequency $f_k$.
The antenna spacing $d$ is chosen on the order of $d=\lambda/2$, where $\lambda$ is the carrier wavelength.

Each antenna provides two switchable feeds, denoted by $\mathrm{R}$ and $\mathrm{L}$.
Those correspond to the two $\pm 45^\circ$ slant-polarized feeds and include the intentional cross-polarization response of the patch design.
Because the receivers behind each antenna can only observe one feed at a time, receivers switch randomly and independently between $\mathrm{R}$ and $\mathrm{L}$.
Measurements are tagged with their \ac{RF} switch state, stored in a short backlog and combined under a quasi-static scene assumption.

The channel estimates for each packet are time-aligned such that the first significant impulse response peak is aligned to a fixed reference tap.
By combining multiple measurements from the backlog, the \ac{SNR} can be improved by averaging, and the relative phases of both feeds of all antennas can be recovered.
Due to \ac{CFO}, each packet has an unknown global phase offset, so an iterative global phase alignment algorithm is needed to combine multiple packets coherently.

The resulting dual-feed snapshot is stacked into the measured $\mathrm{R}/\mathrm{L}$ feed vector
\begin{equation*}
    \mathbf{h}_{m,n,k}^{\mathrm{RL}}
    =
    \begin{bmatrix}
        H_{m,n,k}^{\mathrm{R}} \\
        H_{m,n,k}^{\mathrm{L}}
    \end{bmatrix} \in \mathbb{C}^{2}.
\end{equation*}
For polarization visualization, this feed vector is transformed into separated vertical and horizontal field components by an empirically determined Jones matrix~\cite{jones1941calculus}.
The matrix captures the effective slant-feed response, the intentional cross-polarization terms of the patch, and the physical orientation of antenna $(m,n)$ in the combined array.
With $\mathbf{J}_{m,n}$ denoting this effective feed response, we obtain the corresponding Jones vector in the vertical / horizontal ($\mathrm{V}/\mathrm{H}$) basis, used for all vectors from now on, as
\begin{equation}
    \mathbf{h}_{m,n,k}^{\mathrm{VH}}
    =
    \mathbf{J}_{m,n}^{-1}\mathbf{h}_{m,n,k}^{\mathrm{RL}}.
    \label{eq:jones}
\end{equation}

\subsection{From \acs{CSI} to Angular Spectrum}
For a planar array with half-wavelength spacing, a plane wave arriving from azimuth $\varphi$ and elevation $\vartheta$ causes approximately linear phase progressions~\cite{vantrees2002optimum}
\begin{equation}
    \Psi_{\mathrm{x}} = \pi\cos(\vartheta)\sin(\varphi),
    \qquad
    \Psi_{\mathrm{y}} = \pi\sin(\vartheta)
    \label{eq:angle_to_beamspace}
\end{equation}
over the two array axes.
The pair $(\Psi_{\mathrm{x}},\Psi_{\mathrm{y}})$ is the beamspace coordinate, also interpretable as the phase difference between neighboring antenna elements along the horizontal and vertical array axes.
Not every point in the beamspace corresponds to a physical arrival direction.
We call the subset that maps to real azimuth and elevation angles the \emph{visible region}.
For half-wavelength antenna spacing, this region is the disk
$\Psi_{\mathrm{x}}^2+\Psi_{\mathrm{y}}^2\leq \pi^2$.
We can transform the measured channel into beamspace domain using the two-dimensional Fourier transform
\begin{equation}
    \mathbf{b}_{k}^{\mathrm{VH}}(\Psi_{\mathrm{x}},\Psi_{\mathrm{y}})
    =
    \frac{1}{MN}
    \sum_{m=0}^{M-1}\sum_{n=0}^{N-1}
    \mathbf{h}_{m,n,k}^{\mathrm{VH}}\,
    \mathrm{e}^{-\mathrm{j}(m\Psi_{\mathrm{y}}+n\Psi_{\mathrm{x}})}.
    \label{eq:beamspace}
\end{equation}
In practice,~\eqref{eq:beamspace} is evaluated at a set of discretized beamspace coordinates $(\Psi_{\mathrm{x}},\Psi_{\mathrm{y}})$ by zero-padding the $m,n$ axes of $\mathbf{h}_{m,n,k}^{\mathrm{VH}}$ and applying a two-dimensional \ac{FFT}.
The result is a beam- and frequency-dependent Jones vector
\begin{equation*}
    \mathbf{b}_{k}^{\mathrm{VH}}(\Psi_{\mathrm{x}},\Psi_{\mathrm{y}})
    =
    \begin{bmatrix}
        B_{k}^{\mathrm{V}}(\Psi_{\mathrm{x}},\Psi_{\mathrm{y}}) \\
        B_{k}^{\mathrm{H}}(\Psi_{\mathrm{x}},\Psi_{\mathrm{y}})
    \end{bmatrix} \in \mathbb{C}^2,
\end{equation*}
with $B^{\mathrm{V}}$ and $B^{\mathrm{H}}$ denoting the beamspace channel responses of the vertical and horizontal components, respectively.
The beamspace-domain power is averaged over subcarriers as
\begin{equation*}
    P(\Psi_{\mathrm{x}},\Psi_{\mathrm{y}})
    =
    \frac{1}{K}
    \sum_{k}
    \left\|
        \mathbf{b}_{k}^{\mathrm{VH}}(\Psi_{\mathrm{x}},\Psi_{\mathrm{y}})
    \right\|_2^2.
\end{equation*}
Here, $K$ is the number of subcarriers included in the average.
In the implementation, the spatial \ac{FFT} is accelerated by exploiting time-domain sparsity: We transform the frequency-domain data to the delay domain first and retain only a short window around the synchronized first-arrival tap, for which the beamspace transform is computed.
Afterwards, the beamspace-delay domain data can be transformed back to the frequency domain, which is, overall, still faster than computing the beamspace transform for all subcarriers.

\subsection{Camera-Registered Augmented Reality Overlay}
The augmented-reality display requires a mapping from beamspace to camera pixels.
Using~\eqref{eq:angle_to_beamspace}, a beamspace coordinate is first converted to angles by
\begin{equation}
    \vartheta = \arcsin\left(\frac{\Psi_{\mathrm{y}}}{\pi}\right),
    \qquad
    \varphi = \arcsin\left(\frac{\Psi_{\mathrm{x}}}{\pi\cos(\vartheta)}\right).
    \label{eq:beamspace_to_angle}
\end{equation}
Assuming aligned array and camera boresights, and neglecting the small parallax between camera and array center, these angles are projected into the camera image with a pinhole model using the standard field-of-view normalization of perspective projection~\cite{pharr2017physically}.
For camera field-of-view angles $\Phi_{\mathrm{x}}$ and $\Phi_{\mathrm{y}}$, the normalized image coordinates $(u,v) \in [0,1]^2$ are
\begin{equation}
    u = \frac{1}{2}+\frac{1}{2}\frac{\tan(\varphi)}{\tan(\Phi_{\mathrm{x}}/2)},
    \quad
    v = \frac{1}{2}+\frac{1}{2}\frac{\tan(\vartheta)}{\cos(\varphi)\tan(\Phi_{\mathrm{y}}/2)}.
    \label{eq:camera_projection}
\end{equation}
Fixed azimuth and elevation offsets can be applied to compensate small manual alignment errors between antenna and camera orientations.

\subsection{Delay and Polarization Visualization}
For delay visualization, we assume that each resolved beam with beamspace coordinates $(\Psi_{\mathrm{x}},\Psi_{\mathrm{y}})$ is dominated by a single propagation delay.
The adjacent-subcarrier phase increment then provides a simple and fast relative delay estimate
\begin{equation*}
    \hat{\tau}(\Psi_{\mathrm{x}},\Psi_{\mathrm{y}})
    =
    \frac{
        \arg\bigl(
            \sum_{k}
            \left(\mathbf{b}_{k}^{\mathrm{VH}}(\Psi_{\mathrm{x}},\Psi_{\mathrm{y}})\right)^{\mathsf{H}}
            \mathbf{b}_{k+1}^{\mathrm{VH}}(\Psi_{\mathrm{x}},\Psi_{\mathrm{y}})
        \bigr)
    }{2\pi\Delta f},
\end{equation*}
where $\Delta f$ is the subcarrier spacing.
The relative delays in the scene are mapped to color hue, while brightness remains controlled by beamspace power.

For polarization visualization, we assume that within each beam, the relative vertical and horizontal field components are approximately constant over the observed band,
\begin{equation*}
    \begin{aligned}
    \mathbf{b}_{k}^{\mathrm{VH}}(\Psi_{\mathrm{x}},\Psi_{\mathrm{y}})
    &\approx
    \alpha_k(\Psi_{\mathrm{x}},\Psi_{\mathrm{y}})
    \boldsymbol{\eta}^{\mathrm{VH}}(\Psi_{\mathrm{x}},\Psi_{\mathrm{y}}),\; \left\lVert\boldsymbol{\eta}^{\mathrm{VH}}\right\rVert_2=1.
    \end{aligned}
\end{equation*}
Here, $\alpha_k$ is an arbitrary complex scalar response and $\boldsymbol{\eta}^{\mathrm{VH}}\in\mathbb{C}^2$ is the Jones vector describing polarization.
We want to visualize relative phases across beams $(\Psi_{\mathrm{x}},\Psi_{\mathrm{y}})$, so the otherwise arbitrary phase of $\boldsymbol{\eta}^{\mathrm{VH}}$ is fixed by the gauge $\alpha_{k_0}(\Psi_{\mathrm{x}},\Psi_{\mathrm{y}})\in\mathbb{R}_{+}$, where $k_0$ is the center subcarrier; hence $\boldsymbol{\eta}^{\mathrm{VH}}$ carries the center-subcarrier beam phase.
We estimate $\boldsymbol{\eta}^{\mathrm{VH}}$ as follows: Let $r\in\{\mathrm{V},\mathrm{H}\}$ be the globally stronger polarization component and $\mathcal{N}(\mathbf{x})=\mathbf{x}/\|\mathbf{x}\|_2$. For each beam,
\begin{equation*}
    \begin{aligned}
    \hat{\boldsymbol{\eta}}^{\mathrm{VH}}(\Psi_{\mathrm{x}},\Psi_{\mathrm{y}})
    &=
    \mathcal{N}\left(
        \sum_{k}
        \mathbf{b}_{k}^{\mathrm{VH}}(\Psi_{\mathrm{x}},\Psi_{\mathrm{y}})
        \left(B_{k}^{r}(\Psi_{\mathrm{x}},\Psi_{\mathrm{y}})\right)^*
    \right)\\
    &\quad{}\cdot
    \mathrm{e}^{\mathrm{j}\arg B_{k_0}^{r}(\Psi_{\mathrm{x}},\Psi_{\mathrm{y}})}.
    \end{aligned}
\end{equation*}
For rendering, one common phase rotation is applied to all beams such that
$\sum_{\Psi_{\mathrm{x}},\Psi_{\mathrm{y}}}P(\Psi_{\mathrm{x}},\Psi_{\mathrm{y}})
\hat{\eta}^{\mathrm{V}}(\Psi_{\mathrm{x}},\Psi_{\mathrm{y}})\in\mathbb{R}_{+}$.
This stabilizes the remaining scene-wide animation phase across frames without changing relative phases between beams.
The visualization shows polarization as the strongly slowed-down trajectory traced by the tip of the electric field phasor,
\begin{equation}
    \mathbf{e}^{\mathrm{VH}}(t;\Psi_{\mathrm{x}},\Psi_{\mathrm{y}})
    =
    \Re\left\{
        \hat{\boldsymbol{\eta}}^{\mathrm{VH}}(\Psi_{\mathrm{x}},\Psi_{\mathrm{y}})\mathrm{e}^{\mathrm{j}\omega t}
    \right\},
    \label{eq:polarization_trace}
\end{equation}
where $t$ is the visualization time and $\omega$ is the chosen angular frequency of the animation.
The phasors from \eqref{eq:polarization_trace} are drawn as animated dots (with a trailing trace line) on a regular grid in the camera image.
Vertical, horizontal, slanted, and circular or elliptical polarization states therefore appear as different motion patterns.
Because the visualization preserves relative phases between beams, out-of-phase beams with similar polarization can still be distinguished by their different starting phases.

\subsection{Hardware-Accelerated Rendering}
The rendering pipeline is split between the \ac{CPU} and \ac{GPU}.
The \ac{CPU} computes low-resolution textures containing discretized beamspace power $P(\Psi_{\mathrm{x}},\Psi_{\mathrm{y}})$ and, optionally, delay or polarization information $\hat \tau(\Psi_{\mathrm{x}},\Psi_{\mathrm{y}})$ or $\hat{\boldsymbol{\eta}}^{\mathrm{VH}}(\Psi_{\mathrm{x}},\Psi_{\mathrm{y}})$.
A vertex shader performs the camera-to-beamspace coordinate interpolation according to~\eqref{eq:beamspace_to_angle} and~\eqref{eq:camera_projection}.
A fragment shader blends the resulting texture samples with the camera's video feed and draws the electric field traces for polarization visualization.

\section{Experiments and Results}
\begin{figure}
    \centering
    \includegraphics[width=\columnwidth]{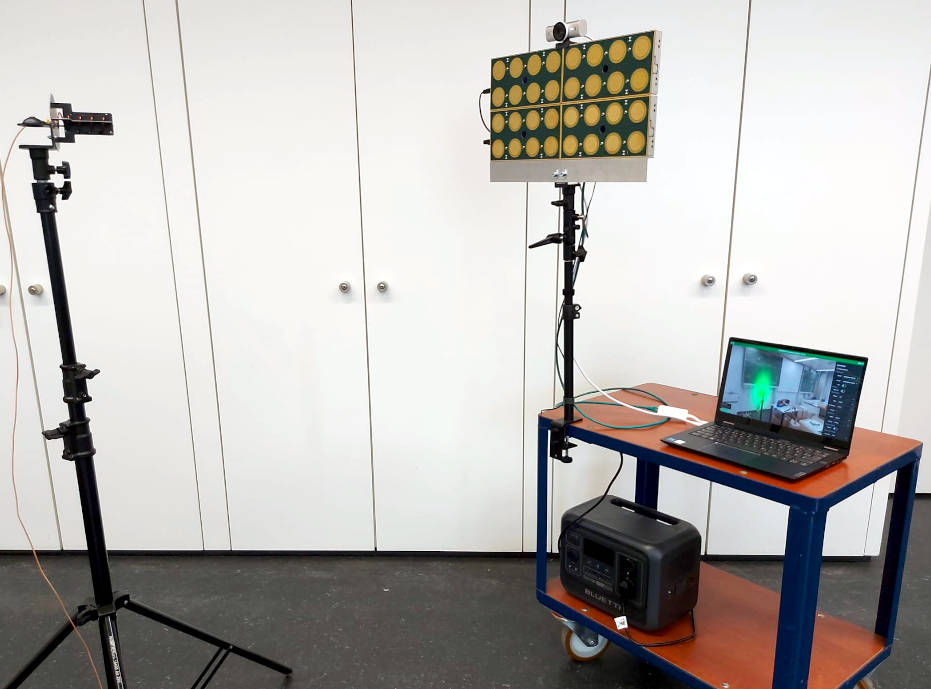}
    \caption{Setup: Four ESPARGOS boards and a webcam, with a total of 32 antenna elements. The \ac{CSI} is streamed to a laptop over Ethernet for real-time processing and visualization.}
    \label{fig:experiment-setup}
\end{figure}

\begin{figure*}
    \centering
    \begin{subfigure}[t]{0.31\textwidth}
        \includegraphics[width=\textwidth]{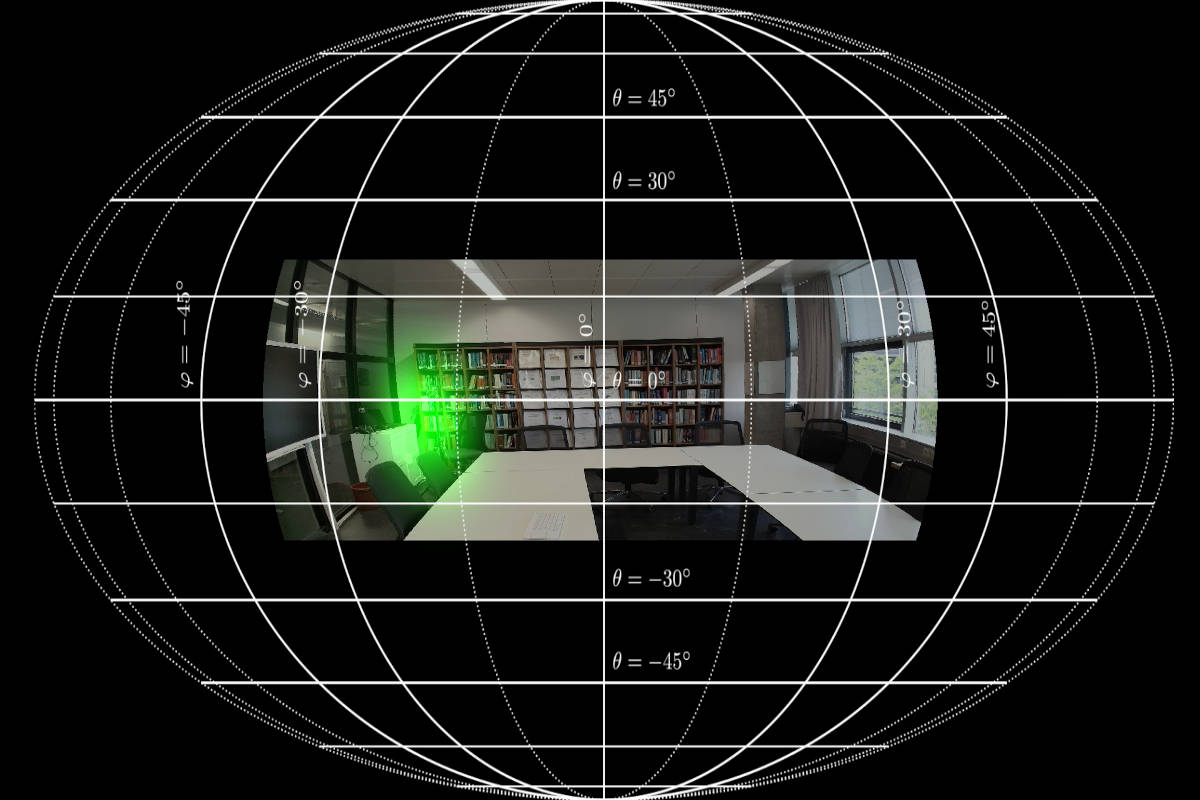}
        \caption{Field of view of a typical webcam compared to the ESPARGOS antenna array's field of view (azimuth / elevation grid), both in beamspace.}
        \label{fig:result-fov}
    \end{subfigure}
    \hspace{0.2cm}
    \begin{subfigure}[t]{0.31\textwidth}
        \includegraphics[width=\textwidth]{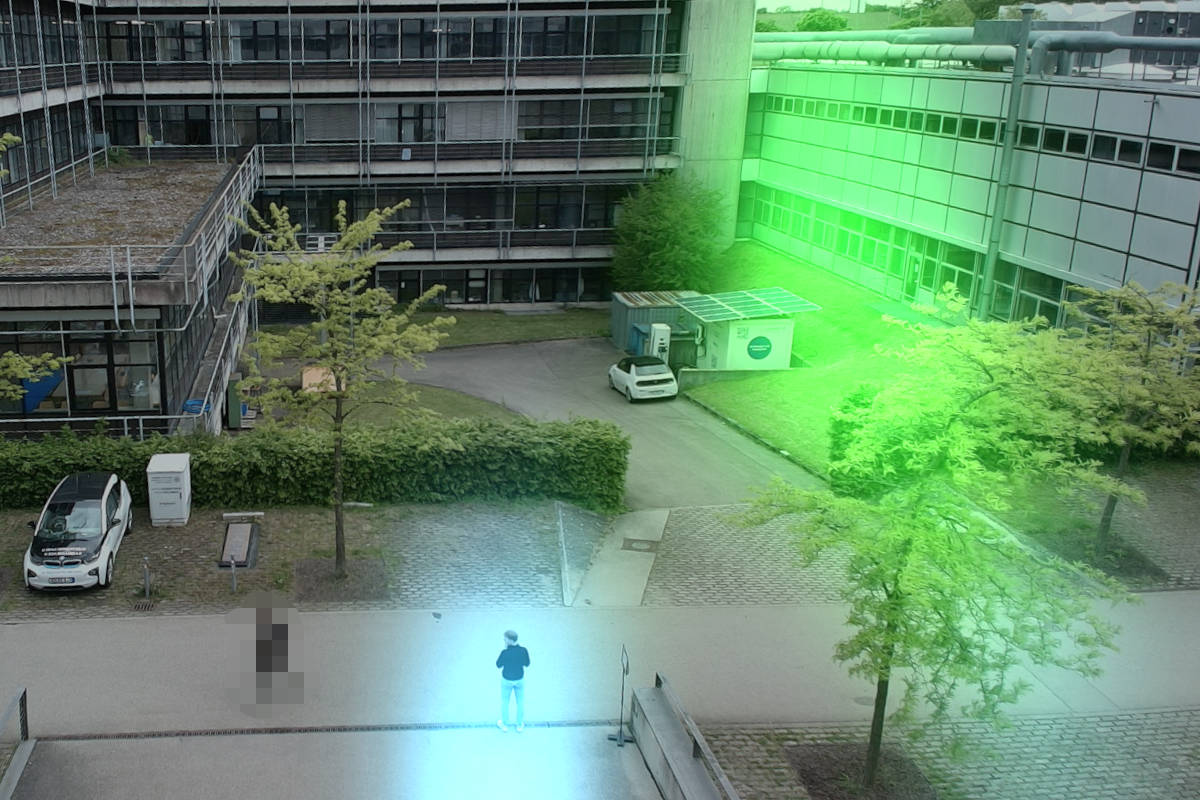}
        \caption{Outdoor paths; the transmitter is held by the person at the bottom. Color shows delay: direct or ground reflection in blue, building-corner reflection in green.}
        \label{fig:result-delay}
    \end{subfigure}
    \hspace{0.2cm}
        \begin{subfigure}[t]{0.31\textwidth}
        \includegraphics[width=\textwidth]{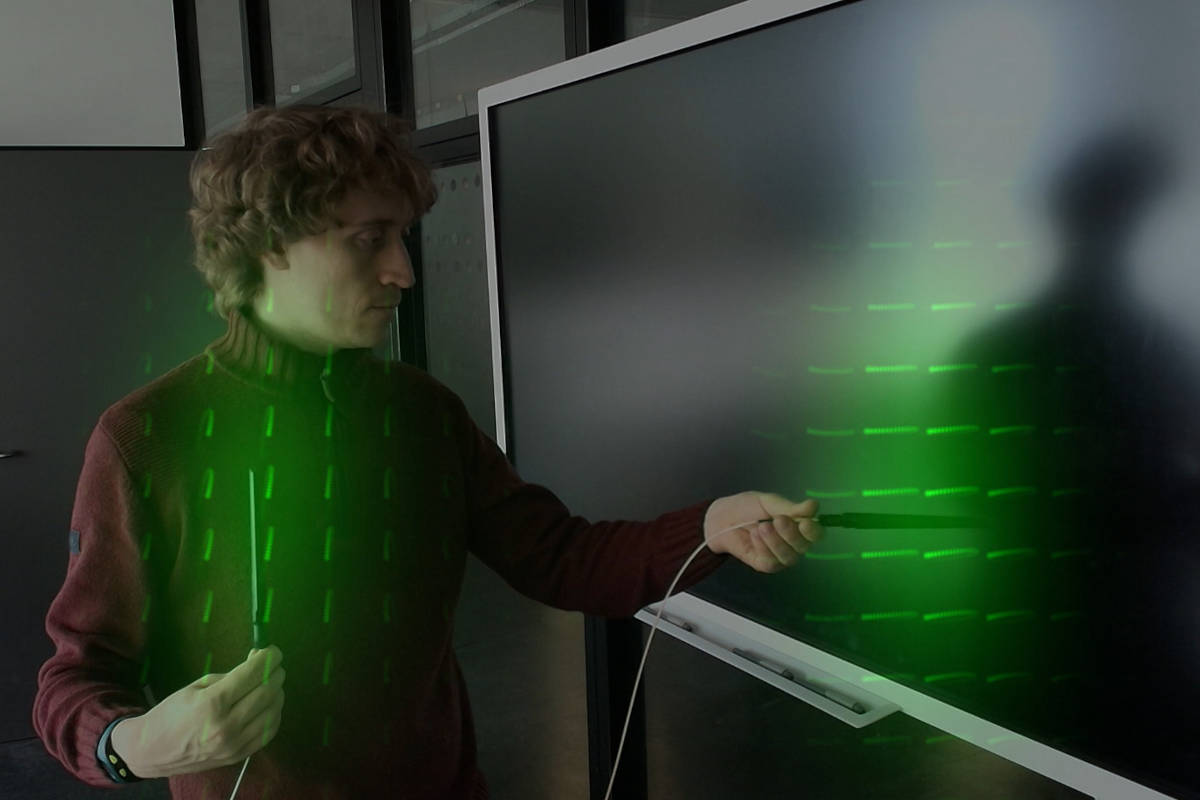}
        \caption{Two dipole antennas transmitting the same signal, but with different polarizations and phase offsets.}
        \label{fig:result-vertical}
    \end{subfigure}

    \vspace{0.2cm}

    \begin{subfigure}[t]{0.31\textwidth}
        \includegraphics[width=\textwidth]{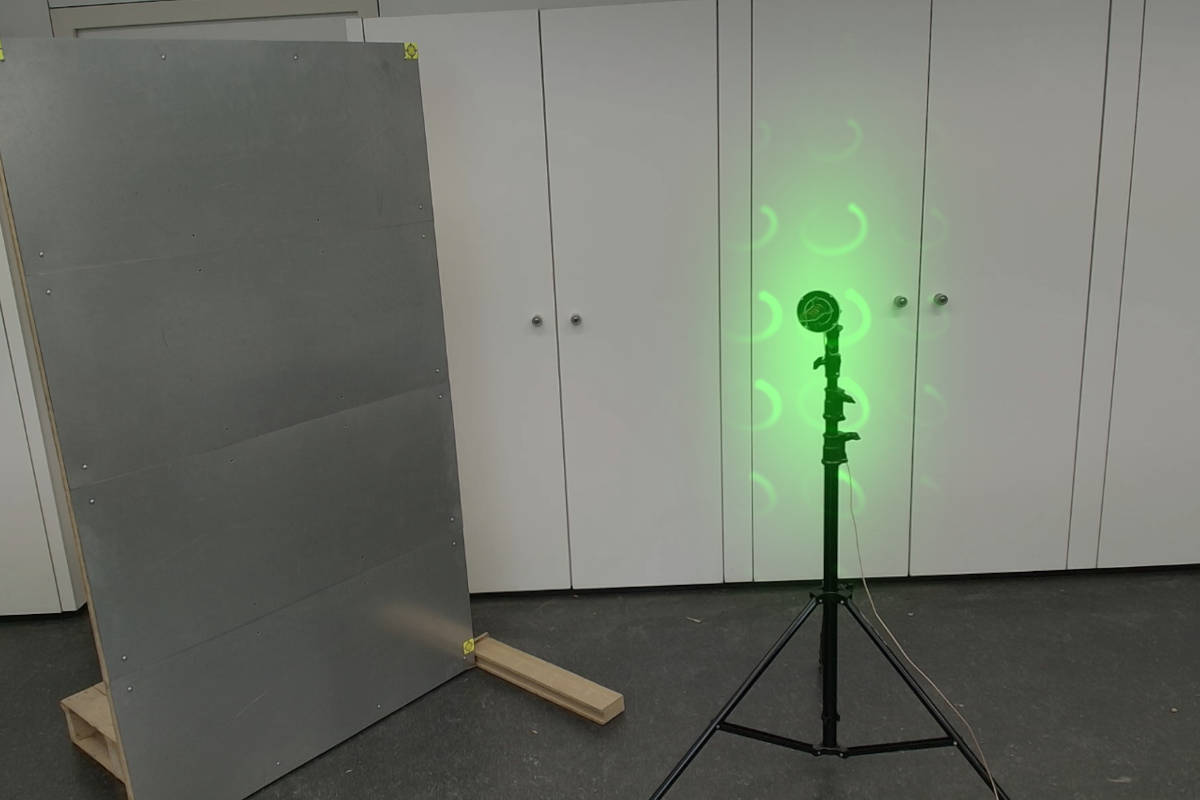}
        \caption{Visualization of an LHCP-polarized signal from a circularly polarized transmit antenna received over a line-of-sight path: dots rotating in a circle, counterclockwise.}
        \label{fig:result-circular-direct}
    \end{subfigure}
    \hspace{0.2cm}
    \begin{subfigure}[t]{0.31\textwidth}
        \includegraphics[width=\textwidth]{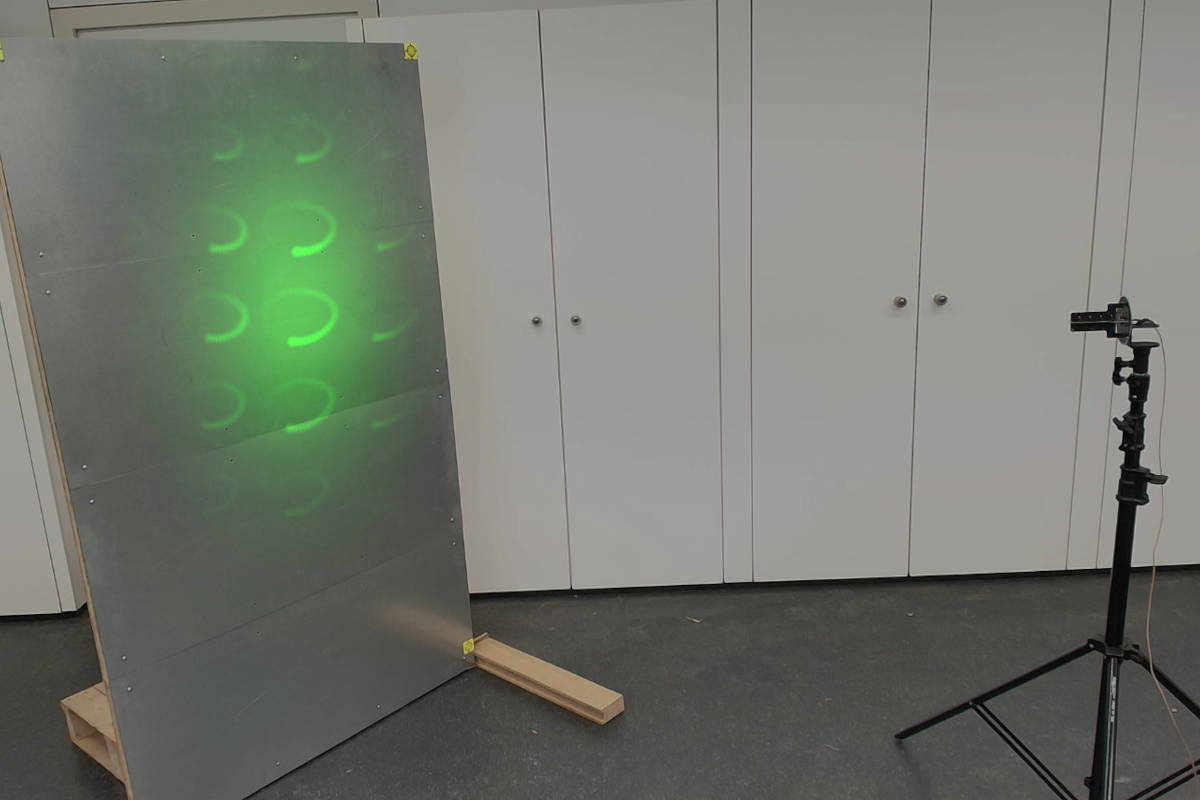}
        \caption{Same circularly polarized transmitter from Fig.~\ref{fig:result-circular-direct}, but now received over a reflection from a metallic wall: dots with clockwise rotation indicate a change from LHCP to RHCP.}
        \label{fig:result-circular-reflected}
    \end{subfigure}
    \hspace{0.2cm}
    \begin{subfigure}[t]{0.31\textwidth}
        \includegraphics[width=\textwidth]{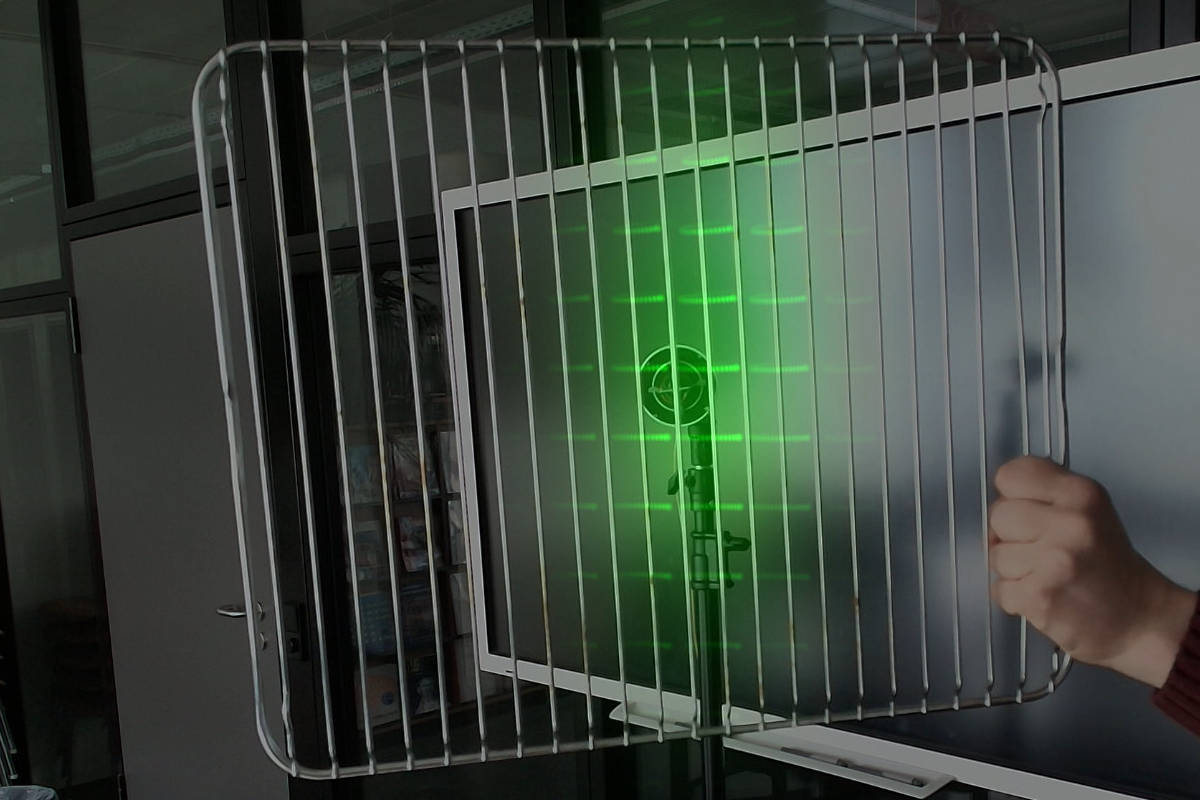}
        \caption{Same circularly polarized transmitter as in Fig.~\ref{fig:result-circular-direct}, but now with a wire rack in the propagation path, acting as a polarization-selective filter.}
        \label{fig:result-filter}
    \end{subfigure}
    \caption{Screenshots from the augmented-reality demo: example visualizations for selected scenarios.}
    \label{fig:results}
\end{figure*}

The experiments use the combined $4\times8$ ESPARGOS array shown in Fig.~\ref{fig:experiment-setup}.
All results are captured and rendered live, with the laptop running the visualization pipeline described in Section~\ref{sec:pipeline}.
In operation, the display is video-like and interactive: the overlay updates continuously at high frame rate.
Since this paper can only show static images, Fig.~\ref{fig:results} presents representative screenshots from the live visualization.

\subsection{Angular Spectrum: Beamspace and Augmented Reality}
Fig.~\ref{fig:result-fov} shows the full visible beamspace region of the array together with the smaller region covered by the optical camera.
This illustrates that the array observes a wider angular field of view than the camera, while the augmented-reality overlay uses the camera-covered subset to place the measured angular spectrum on the optical image.

\subsection{Visualizing Delay}
Fig.~\ref{fig:result-delay} illustrates that ESPARGOS separates multipath components in both angle and delay: the direct and ground-reflected components appear earliest, while the angularly separated building-corner reflection appears later.

\subsection{Visualizing Polarization}
Figs.~\ref{fig:result-vertical}--\ref{fig:result-filter} show that ESPARGOS can separate propagation components by angle and polarization at the same time.
In Fig.~\ref{fig:result-vertical}, two dipoles transmitting the same signal are resolved with different polarization and phase.
Figs.~\ref{fig:result-circular-direct} and~\ref{fig:result-circular-reflected} show a circularly polarized signal whose rotation changes between the line-of-sight and reflected path, while Fig.~\ref{fig:result-filter} shows a wire rack acting as a polarization-selective object.

\section{Summary and Outlook}
This paper presented a camera-registered ESPARGOS visualization pipeline that overlays measured beamspace power, relative delay, and polarization onto the optical scene.
Our hardware and software make angular propagation, multipath, and field-orientation effects visible in the physical coordinate frame.
In the future, we plan to extend our system with a more compact, handheld hardware device and to explore the radar sensing (phase-coherent transmission) capabilities of ESPARGOS with a similar visualization approach.

\IfFileExists{IEEEtran.bst}{\bibliographystyle{IEEEtran}}{\bibliographystyle{unsrt}}
\bibliography{references}

\end{document}